\documentclass[aip,apl,reprint,twocolumn,superscriptaddress,noshowpacs,notitlepage,longbibliography,10pt,citeautoscript]{revtex4-2}%

\usepackage{graphicx,bm,times}
\graphicspath{ {./figures/main/} }
\usepackage{amsmath}
\usepackage{amsfonts}
\usepackage{amssymb}
\usepackage{mathtools}
\usepackage{color}
\usepackage{hyperref}
\hypersetup{
	colorlinks = true,
	allcolors = {blue}
}

\begin{document}

\title{Substrate pre-sputtering for layer-by-layer van der Waals epitaxy of 2D materials}

\author{A.~Rajan}
\email{ar289@st-andrews.ac.uk}
\affiliation{SUPA, School of Physics and Astronomy, University of St Andrews, St Andrews KY16 9SS, UK}

\author{M.~Ramirez}
\affiliation{SUPA, School of Physics and Astronomy, University of St Andrews, St Andrews KY16 9SS, UK}

\author{N.~Kushwaha}
\affiliation{SUPA, School of Physics and Astronomy, University of St Andrews, St Andrews KY16 9SS, UK}
\affiliation{STFC Central Laser Facility, Research Complex at Harwell, Harwell Campus, Didcot OX11 0QX, United Kingdom}

\author{S.~Buchberger}
\author{M.~McLaren}
\affiliation{SUPA, School of Physics and Astronomy, University of St Andrews, St Andrews KY16 9SS, UK}

\author{P.D.C.~King}
\email{pdk6@st-andrews.ac.uk}
\affiliation{SUPA, School of Physics and Astronomy, University of St Andrews, St Andrews KY16 9SS, UK}

\date{\today}

\begin{abstract} 
Two-dimensional transition metal chalcogenides, with their atomically layered structure, favourable electronic and mechanical properties, and often strong spin-orbit coupling, are ideal systems for fundamental studies and for applications ranging from spintronics to optoelectronics. Their bottom-up synthesis via epitaxial techniques such as molecular-beam epitaxy (MBE) has, however, proved challenging. Here, we develop a simple substrate pre-treatment process utilising exposure to a low-energy noble gas plasma. We show how this dramatically enhances nucleation of an MBE-grown epilayer atop, and through this, realise a true layer-by-layer growth mode. We further demonstrate the possibility of tuning the resulting growth dynamics via control of the species and dose of the plasma exposure.
\end{abstract}

\maketitle

The large family~\cite{chhowalla2013chemistry, manzeli20172d} of MX$_2$ transition-metal dichalcogenides (TMDs, M=transition metal, X=\{S,Se,Te\}) and related self-intercalated M$_{1+\delta}$X$_2$ sister compounds have found widespread application, including as 2D spin-orbit coupled semiconductors~\cite{mak2010mos2, mak2012control, Xiao2012mos2, riley_direct_2014}, magnets~\cite{gong_discovery_2017,huang_layer-dependent_2017,Wang2018CGTmagnetism, Verzhbitskiy2020magnetism}, superconductors~\cite{ugeda2016characterization,Xi2016islingnbse2,Bawden2016spinvalley}, charge density-wave systems~\cite{ugeda2016characterization, feng2018electronic} and topological insulators~\cite{soluyanov_type-ii_2015,bahramy_ubiquitous_2018}. The layered nature of the parent compounds allows thinning the host material down to individual MX$_2$ monolayers, with their materials properties typically found to depend sensitively on thickness. Examples include: the crossover from an indirect to a direct band gap semiconductor in MoS$_2$ as a result of quantum confinement~\cite{mak2010mos2, splendiano2010mos2}; a marked increase in exciton binding energy due to the reduced dielectric screening~\cite{chernikov_exciton_2014,ugeda_giant_2014,Raja2017monomx2,buchberger2025persistencechargeorderinginstability}; the emergence of an Ising superconducting state due to the loss of inversion symmetry in the monolayer limit~\cite{Xi2016islingnbse2, Bawden2016spinvalley, xing2017islingnbse2}; and the realisation of distinct topological and magnetic states depending on the layer number~\cite{huang_layer-dependent_2017,cucchi_microfocus_2019,kushwaha2025ferromagnetic}. This demands the ability to control the materials thickness with digital monolayer precision.

The most common way to achieve such atomic-scale thickness control is via the mechanical exfoliation of flakes from a bulk crystal, using the famous scotch-tape-type methods~\cite{novoselov2005exfoliation, chhowalla2013chemistry}. Sophisticated transfer methods have now been developed to allow the fabrication of van der Waals heterostructures and devices from the exfoliated flakes~\cite{Wang2023heterostructures}. Typically, however, multiple flakes of different thickness are obtained from a single exfoliation, with the lateral spatial extent of monolayer regions often limited. Challenges with contamination~\cite{Kretinin2014graphenedevice, haigh2012}, air stability~\cite{fan2016degradation}, and scaleability~\cite{LIU2021exfoliation} all motivate the development of direct bottom-up synthesis techniques for the fabrication of TMD and other TM-chalcogenide (TMC)-based 2D materials. However, the requirement for digital thickness control has remained extremely challenging. Techniques such as molecular beam epitaxy (MBE) -- well known for the precise layer-by-layer fabrication of conventional semiconductors -- have suffered from poor growth uniformity. Apart from a few pioneering studies~\cite{pacuski2020narrow}, it has not typically been possible to grow monolayer islands larger than ca.\ 1 micrometer in lateral extent using this technique, before the onset of bilayer regions which leads to significant thickness inhomogeneity~\cite{rajan2020morphology}.

Recently, however, we have found a general route to overcome this problem. Exposing the growth surface to a small ion flux of a sacrificial species such as Ge or Ag, we found a significant enhancement in nucleation of the epilayer in monolayer TMC growth~\cite{Adv.mater.2024}. This led to a dramatic enhancement in the growth rate and coverage, enabling the formation of large-area epitaxial monolayers and van der Waals heterostructures. We found no evidence for incorporation of the sacrificial species into the film -- indeed obtaining enhanced quasiparticle lifetimes in our samples as compared to ones grown via conventional MBE. Nonetheless, to completely eliminate the possibility of impurity incorporation, it would be advantageous to avoid the use of sacrificial species during the growth at all. Moreover, our fabrication procedure required the use of an electron-beam evaporator during the growth, which is not routinely available in all MBE systems. To this end, here we report an alternative method to achieve a similar result, requiring only the pre-treatment of the growth substrate using a simple, quick, and readily available, ultra-high vacuum (UHV) sputtering technique. We show how this can be used to control nucleation on the substrate with a high degree of flexibility, and demonstrate how this can be used to achieve a desired layer-by-layer growth mode of 2D TMDs and TMCs.

\begin{figure*}
    \centering
    \includegraphics[width=\textwidth]{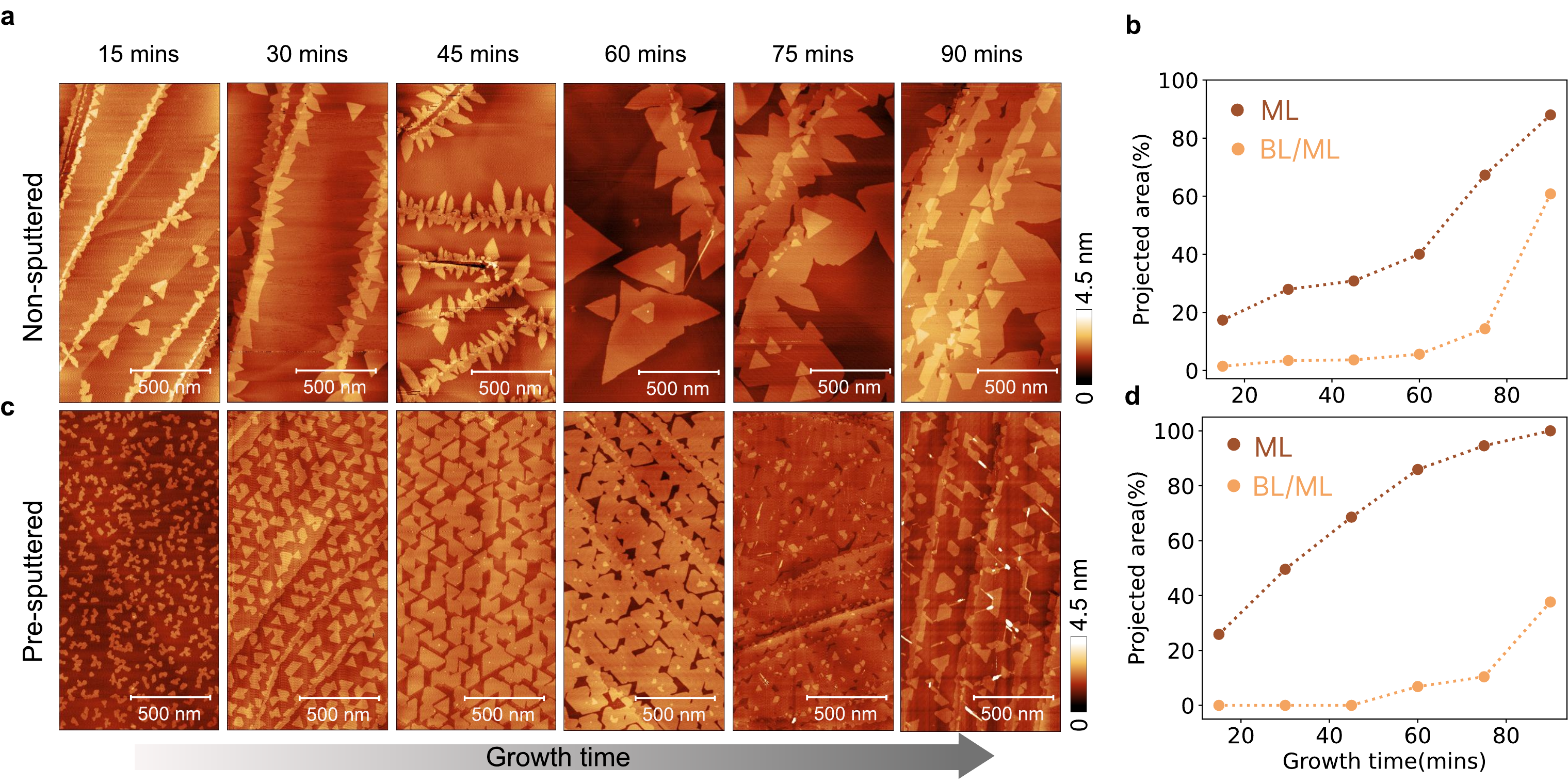}
    \caption{\label{f:fig1} (a) AFM topographies of TiSe\textsubscript{2} samples grown on un-treated HOPG substrates as a function of increasing growth times. (b) Corresponding mono- and bilayer coverage extracted from analysis of our AFM measurements. (c,d) Corresponding measurements for samples where the substrate has been  pre-sputtered using He for 10s. These samples show a striking enhancement in the nucleation, growth rate and suppression of bilayers as compared to those grown on pristine substrates.}
\end{figure*}

All growths reported here were performed using a DCA R450 MBE system, dedicated to the growth of chalcogenide materials. Ti and Cr were supplied from high-temperature effusion cells, containing 3N5 pure Ti and 4N pure Cr. A valved cracker cell was used for the supply of a 5N pure Se flux, with the cracker zone held at 500~$^\circ$C during the growth. 6N pure Te was evaporated from a home-built Knudsen cell. Fluxes from the cells were measured in beam-equivalent pressure (BEP), by positioning a retractable ion gauge as a beam flux monitor in front of the substrate, just before the growth.

Highly oriented pyrolytic graphite (HOPG) substrates were used throughout. Fresh HOPG surfaces were exfoliated in atmosphere before rapidly transferring the samples into a vacuum load lock. The substrates were then degassed at 200~$^\circ$C in the load lock overnight. Except for the reference samples shown in Fig.~\ref{f:fig1}(a), the substrates were then transferred to a UHV preparation chamber, where they were lightly sputtered using a Specs IQE 11/35 sputter gun. A He (Ar) plasma was utilised, with a partial pressure of $7.7\times10^{-7}$ ($5.5\times10^{-6}$)~mbar as measured by an ionisation gauge in the UHV chamber. Following sputtering, the substrates were transferred to the growth chamber. All substrates were subsequently annealed at between 700~$^\circ$C and 800~$^\circ$C for 30 minutes before cooling to the growth temperature and starting the growth. After growth, the samples were removed from the vacuum and immediately transferred to an Ar glovebox. Their morphology was probed using a Park Systems NX10 Atomic Force Microscope (AFM) placed within the glovebox, utilising a non-contact measurement mode.

Figure~\ref{f:fig1}(a) summarises the known growth mode~\cite{rajan2020morphology,peng_molecular_2015} of a TiSe$_2$ layer using a typical MBE procedure, with co-evaporation of Ti and Se directly onto the freshly exfoliated and degassed substrate surface. Throughout the growth, we supply at least two orders of magnitude higher chalcogen flux than the metal flux, due to the high vapour pressures and low sticking coefficients of the former. The growth is therefore dominated by the sluggish kinetics of the metal adatoms. Nucleation in the early stages of the growth is limited, and preferentially occurs at step edges of the substrate and other substrate defects. With increasing deposition time, growth extends from these nucleated regions. Well before this extends to a complete monolayer coverage, however, a significant onset of bilayer formation is visible (Fig.\ref{f:fig1}(b)).

As discussed above, our previous work indicated that this monolayer morphology can be significantly enhanced by the co-evaporation of a small Ge or Ag ion flux during the growth, which we attributed to the creation of nucleation sites at the substrate surface by the incident excited ions.~\cite{Adv.mater.2024} Here, we attempted to achieve the same enhancement in substrate nucleation without requiring the co-evaporation of an impurity species during the growth. 

\begin{figure*}
    \centering
    \includegraphics[width=\textwidth]{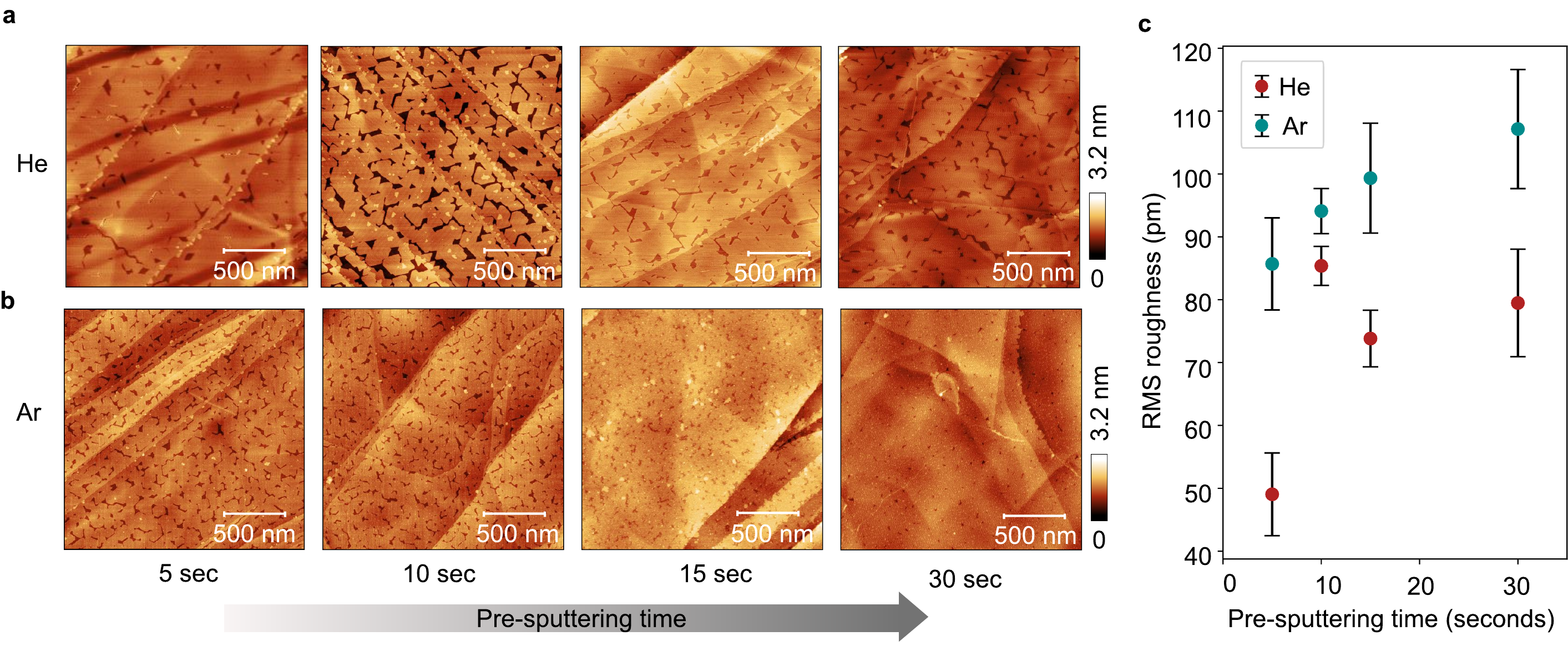}
    \caption{\label{f:fig2} (a) AFM topographies of TiSe\textsubscript{2} samples grown for 60 minutes on HOPG substrates pre-sputtered with He ions for between 5 and 30 seconds as indicated. (b) Corresponding measurements for growth on Ar pre-sputtered substrates. (c) Root mean square (RMS) roughness of the fabricated films, extracted as the average RMS roughness from five $78\times{78}$~nm regions taken from separate locations of the measured AFM topography atop the monolayer island regions (see Supplementary Fig.~S1).}
\end{figure*}

To this end, we pre-sputtered the substrates with ions generated using a noble gas plasma. We show the resulting fabricated samples in Figure \ref{f:fig1}(c). Here, we have pre-exposed the substrate to excited ions generated by a conventional ion source used for sputter-cleaning in UHV. Here, however, we have limited the energy to 200~V -- the lowest voltage in which we could sustain a stable plasma using our commercial UHV ion source -- well below the values typically used for sputter cleaning. Initially, we also investigate the use of very light He ions, and furthermore limit the exposure time to only 10~s. Nonetheless, we already observe a marked difference in the subsequent epilayer growth on these substrates. After only 15 minutes of growth, the pre-sputtered substrate shows a large number of independent islands across the entire growth surface. This is in stark contrast to growth on the pristine substrate, where the deposition almost exclusively occurs at substrate step edges. This indicates that substrate defects are introduced by the pre-sputtering, which act as effective nucleation centres to seed subsequent growth of the TiSe$_2$ layer. Similar to the case for Ge-assisted growth reported previously~\cite{Adv.mater.2024}, once these nucleation sites are established, they can rapidly expand, capturing diffusing metal adatoms which can then participate in edge diffusion. This leads to the formation of well-defined compact triangular islands after 30 minutes of growth. These steadily gain size and start to merge into a large nearly-continuous monolayer at a growth time of around 60 minutes. From our quantitative analysis (Fig.~\ref{f:fig1}(b,d)), we find that the growth rate is enhanced by more than a factor of two: growth on the pre-sputtered substrate has $\approx\!80$\% monolayer coverage after 60 minutes, while growth on the untreated substrate under otherwise identical conditions yields only $\approx40$\% monolayer coverage. With further increase in growth time on the pre-sputtered substrate, the remaining monolayer gaps are filled in to form a continuous monolayer coverage, with the onset of significant bilayer formation only starting to onset at this point (Fig.~\ref{f:fig1}(d)). This is in contrast to the untreated substrate, where the enhancement of mono- and bilayer coverage proceed in parallel after ca. 60 minutes of growth, with a small amount of trilayer coverage already starting to form by the longest growth time shown here.

The growth on the pre-sputtered substrate is therefore reflective of a layer-by-layer growth mode, of the form desired to allow the fabrication of large-area uniform layers of 2D materials by MBE. This is achieved without requiring the co-evaporation of any impurity species during the growth, and so provides an ultra-clean method for achieving digital thickness control of these materials. In the following, we focus on the tunability, generality, and optimisation of this pre-sputtering process.

We show in Fig.~\ref{f:fig2}(a) the impact of varying the time of the He pre-sputtering process. In all cases, a high monolayer coverage is evident, with negligible bilayer formation. There is only a marginal increase in the growth rate between 5 sec and 30 sec of He pre-sputtering. This indicates that there are already sufficient nucleation sites here that the growth rate is near saturated. Nonetheless, the underlying islands, which merge to form the continuous monolayer, are clearly larger and with lower density in the 5~s pre-sputtered film as compared to the longer sputtering times. This leads to a lower density of grain boundaries in the fabricated films, and an excellent root-mean-square (RMS) surface roughness of ${49}\pm{6}$~pm as measured from atop the monolayer islands (Fig.~\ref{f:fig2}(c)). We thus conclude that a pre-sputtering of 5~s here appears to provide a rather optimal surface morphology for the growth of monolayer TiSe$_2$.

He is not routinely used as a source gas in UHV sputtering systems, however. We have therefore also investigated the pre-sputtering of the growth substrate using the more common, but heavier, Ar ions. We show the results of these in Fig.~\ref{f:fig2}(b). Similar to the case of He pre-sputtering, a high monolayer coverage and the suppression of bilayer formation is clearly evident, indicating general efficacy of the approach. However, with increasing Ar sputtering times, a decrease in the size of the individual islands which have merged to form this nearly-continuous monolayer can be observed. This points to a significant increase in the density of nucleation sites here, as compared to our He pre-sputtered films. While this maintains a high-coverage monolayer, it also leads to a higher grain boundary density where islands nucleated at different sites merge. This is evident as an increased surface roughness of the resulting films, both as compared to our He pre-sputtered films, and as a function of increasing pre-sputtering time (Fig.~\ref{f:fig2}(c)). Moreover, close inspection of the measured AFM topographies reveals the formation of pinpoint clusters forming, whose density increases with increasing sputtering time. This suggests the formation of a more extended defect being formed in the 2D material around the initial nucleation site.

\begin{figure}
    \centering
    \includegraphics[width=\columnwidth]{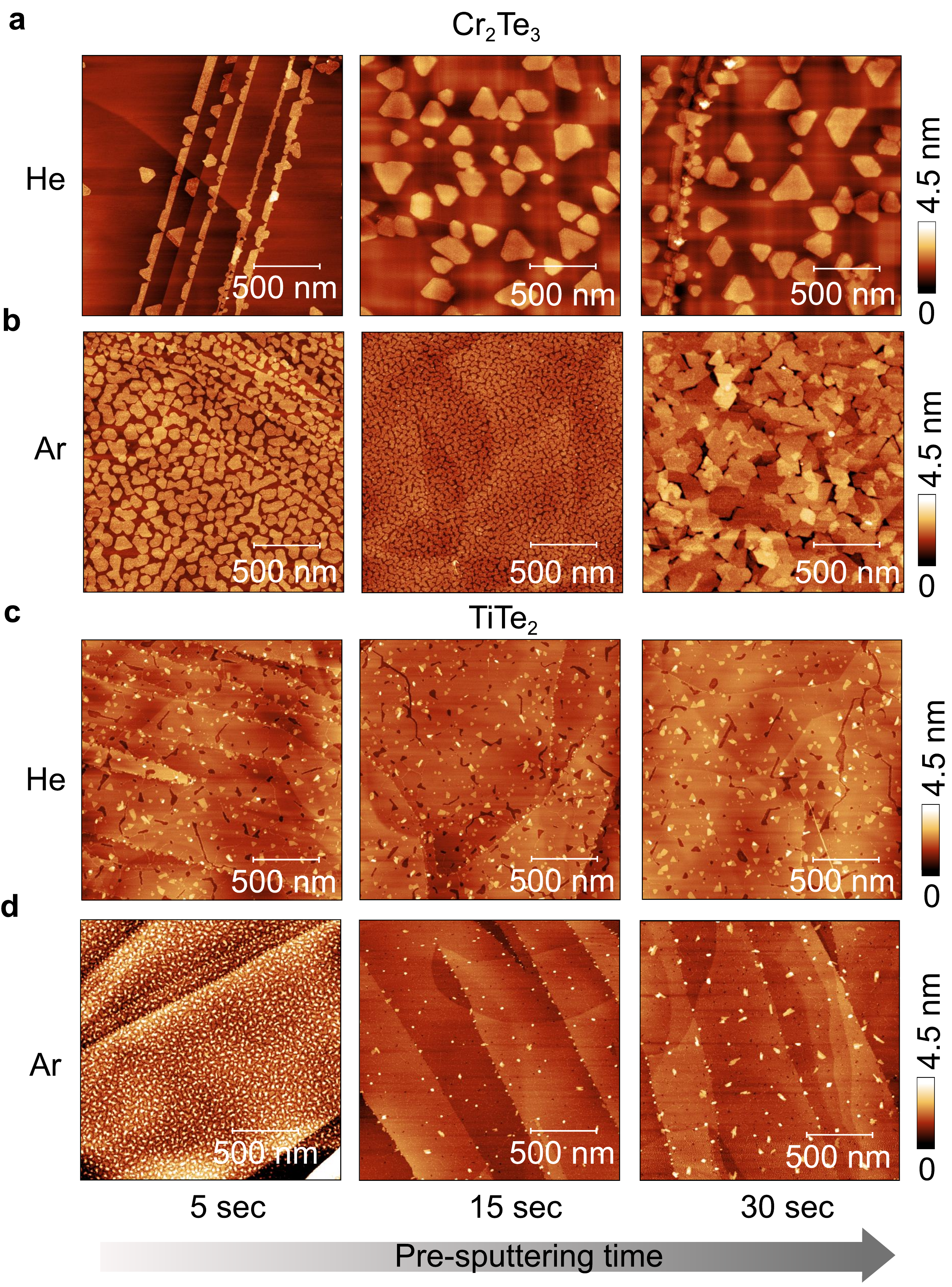}
    \caption{\label{f:fig3} (a,b) AFM topographies from Cr$_2$Te$_3$ samples after 1 hour of growth on HOPG substrates pre-sputtered with (a) He and (b) Ar for 5, 15 and 30~s. These show a stark difference in the number of nucleation sites and the obtained monolayer coverage. (c,d) Equivalent measurements for TiTe$_2$ growth.}
\end{figure}

We therefore conclude that He pre-sputtering is the most effective route for the growth of high-quality TiSe$_2$ monolayers. However, this appears to depend on the specific monolayer compound being fabricated. We show in Fig.~\ref{f:fig3}(a,b) the difference in growth morphologies of Cr$_2$Te$_3$ grown on both He and Ar pre-sputtered substrates. Here, it is evident that the optimal He pre-sputtering conditions as for TiSe$_2$ growth are ineffective for the fabrication of Cr$_2$Te$_3$ (Fig.~\ref{f:fig3}(a)). Minimal nucleation is found away from step edges for the 5~s pre-sputtered case, with no significant differences found to growth on an untreated substrate~\cite{Adv.mater.2024, kushwaha2025ferromagnetic}. The coverage of nucleated islands can be increased by increasing the pre-sputtering time, but the coverage remains relatively low. In contrast, Ar pre-sputtering (Fig.~\ref{f:fig3}(b)) provides a more effective tuning of the growth dynamics. At longer (30~s) pre-sputtering time, too much nucleation results, leading to multilayer Cr$_2$Te$_3$ growth. 15~s pre-sputtering yields high-coverage monolayer growth, but the resulting islands are very small with non-uniform morphologies, and their density very high, leading to a significant number of grain boundaries. The number of nucleation sites can be decreased by lowering the pre-sputtering time, with a uniform coverage of more regular-shaped islands found for a pre-sputtering of 5~s. Together, this suggests that establishing a nucleated island for the growth of Cr$_2$Te$_3$ requires a stronger defect potential on the substrate. This favours the use of the larger Ar atoms, which will create a more significant defect cascade when incident on the substrate. However, to achieve a uniform layer growth and lower grain boundary density, a smaller number of these defects are required, necessitating a low pre-sputtering time.

These results not only point to the possibility to optimise the substrate pre-sputtering for growth of the desired compound, but also highlight a significant material specificity of the growth dynamics and resulting required nucleation landscape. To examine this further, and to investigate whether it is the different anion that leads to the dramatically different behaviour between the two compounds investigated above, we have also studied the sensitivity of TiTe$_2$ growth to our pre-sputtering methodology. Fig.~\ref{f:fig3}(c,d) shows the resulting AFM topographies. Interestingly, here, the 15 and 30~s Ar pre-sputtering appears to lead to a complete monolayer coverage of TiTe$_2$. However, the surface roughness as extracted from our AFM measurements is substantially higher than for TiSe$_2$ growth with the same substrate pre-sputtering conditions ($145\pm10$~pm vs.\ $107\pm9$~pm for TiTe$_2$ and TiSe$_2$ samples grown after 30~s of Ar pre-sputtering, respectively). The reason for this is evident from the 5~s Ar pre-sputtering case: nucleation appears very effective for TiTe$_2$ here, leading to an extremely high density of very small growth islands. While these can still merge to form continuous monolayers (Fig.~\ref{f:fig3}(d)), the high density of grain boundaries leads to a rougher surface, as for the most heavily Ar pre-sputtered TiSe$_2$ growths (Fig.~\ref{f:fig2}(b)). In contrast, here, He pre-sputtering is again highly effective: larger grains form and merge to form near-continuous monolayers with only minimal bilayer formation, and smooth and well-ordered films result with an RMS roughness of only $74\pm10$~pm for the 5~s He pre-sputtered film shown in Fig.~\ref{f:fig3}(c)).

Interestingly, this indicates that the larger anion does have a significant effect on the growth dynamics (cf. Fig.~\ref{f:fig3}(c,d) and Fig.~\ref{f:fig2}). This therefore cannot explain the markedly different growth mode of Cr$_2$Te$_3$. Rather, we speculate that the latter arises as a consequence of the complex growth phase window for Cr-Te growths, with multiple competing metastable states present~\cite{kushwaha2025ferromagnetic}. Nonetheless, our measurements here show that uniform growth of these phases can still be achieved using our simple pre-sputtering approach, if the gas species and pre-sputtering time is appropriately optimised. Indeed, compared to our previous work on Ge-assisted MBE growth of 2D materials~\cite{Adv.mater.2024}, a significant advantage here -- besides the simplicity of the substrate pre-preparation -- is the wide tuneability that can be achieved by changing the gas species, sputtering time, and sputtering voltage. This can be expected to allow optimising the substrate defect landscape for the growth material of interest, allowing the layer-by-layer growth of a wide array of 2D TMCs by MBE. The lack of the use of an impurity species during the growth precludes any unintended incorporation into the growing film. Moreover, the fact that the defect creation occurs purely as a pre-processing step means that unintended defect formation within the growing film itself will be minimised, facilitating effective multi-layer growths. The simplicity and speed of the pre-processing step required here furthermore lends itself to industrial processing, promising to establish MBE as a scaleable route for the fabrication of large-area ultra-pure TMCs.

\section*{Acknowledgements}
We gratefully acknowledge support from the UK Engineering and Physical Sciences Research Council (Grant No.~EP/X015556/1). For the purpose of open access, the authors have applied a Creative Commons Attribution (CC BY) licence to any Author Accepted Manuscript version arising. The research data supporting this publication can be accessed at [[DOI TO BE INSERTED]].

\bibliography{Reference}

\end{document}